\documentstyle[aaspp4,11pt]{article}

\newcommand{\be}{\begin{eqnarray}}
\newcommand{\ee}{\end{eqnarray}}

\def\etal{{\it et al.}}
\def\simless{\mathbin{\lower 3pt\hbox
   {$\rlap{\raise 5pt\hbox{$\char'074$}}\mathchar"7218$}}} 
\def\simgreat{\mathbin{\lower 3pt\hbox
   {$\rlap{\raise 5pt\hbox{$\char'076$}}\mathchar"7218$}}} 

\def\FWHM{{\theta_{\rm FWHM}}}

\def\eps02{{{\epsilon_{-2}}}}

\def\rns{{R_{\rm NS}}}
\def\kms{{km s$^{-1}$}}

\def\vpperp{{{V_p}_{\perp}}}

\def\dnud{{\Delta\nu_{\rm d}}}

\def\dtd{{\Delta t_{\rm d}}}

\def\Thvec{{ \mbox{\boldmath $\Theta$} }}

\def\thmax{{\theta_{\rm max}}}

\def\ld{{\ell_d}}

\def\dphi{{D_{\phi}}}

\def\vp{{V_{\rm p}}}

\def\vpvec{{\bf V_{\rm p}}}
\def\vobsvec{{\bf V_{\rm obs}}}
\def\vismvec{{\bf V_{\rm m}}}

\def\veffvec{{\bf V_{\rm eff}}}

\def\ds{{D_s}}				

\def\c1u{{C_{1,u}}}
\def\c1Kolu{{C_{1,5/3,u}}}

\def\half{{\frac{1}{2}}}

\def\mathbf{{\bf}}
\def\third{{\frac{1}{3}}}

\def\boxit#1{\vbox{\hrule\hbox{\vrule\kern3pt\vbox{\kern3pt#1\kern3pt}
    \kern3pt\vrule}\hrule}}

\def\vareps{{\varepsilon}}

\def\Gambar{{\overline{\Gamma} }}

\def\bvec{{\bf b}}
\def\beffvec{{\bf b_{\rm eff}}}

\def\rsvec{{\bf r_s}}
\def\rs{{r_s}}
\def\drsvec{{\bf \delta r_s}}
\def\drs{{\delta r_s}}

\def\drsiso{{\delta r_{s,iso}}}
\def\rvec{{\bf r}}

\def\Iave{{\langle I \rangle}}

\def\Niave{{\langle N_i \rangle}}
\def\Njave{{\langle N_j \rangle}}

\def\G2ave{{\langle G^2 \rangle }}
\def\sigC{{\sigma_C}}
\def\sigX{{\sigma_X}}

\def\fgam{{f_{\gamma}}}
\def\fGambar{{f_{\Gambar}}}
\def\niss{{N_{\rm ISS}}}
\def\Gbar{{{G}}}

\def\Gambarmag{{\vert \Gambar \vert}}

\def\misssq{{m_{\rm ISS}^2}}
\def\misssqinv{{m^{-2}_{\rm ISS}}}

\def\taubar{{\overline{\tau}}}

\def\Qiss{{Q_{\rm ISS}}}

\def\gamg{{\gamma_{\rm g}}}
\def\gamG{{\gamma_{\rm G}}}

\def\rem{{r_{em} }}

\def\epsem{{\epsilon_{em}}}

\def\rlc{{r_{\rm LC}}}

\def\rbar{{\overline r}}
\def\Drmag{{\Delta r_{em}}}
\def\Drmagperp{{\Delta r_{em,\perp}}}

\def\hatmiss2{{ {\hat m}_{\rm ISS}^2}}
\def\hatmgam2{{ {\hat m}_{\rm \Gamma}^2}}

\def\sigm2{{\sigma_{m^2}}}
\def\SNR{{\rm SNR}}

\def\GI{{G97}}
\def\GII{{G00}}

\tighten
\singlespace

\begin{document}

\title{Interstellar Seeing. II.  The Case of the Vela Pulsar:   Source Unresolved }
\author{J. M.  Cordes}
\affil{Astronomy Department and NAIC, Cornell University, Ithaca, NY 14853
cordes@spacenet.tn.cornell.edu}

\vskip 0.1truein
\centerline{\bf submitted to {\it Astrophysical Journal} \today}

\begin{abstract}

I use a method based on interstellar scintillations for 
discerning information about source sizes on scales less than
one micro-arc sec.  I use a comprehensive model for a pulsar signal,
scintillated amplitude modulated noise, that 
includes source fluctuations and noise statistics. 
The method takes into account time-frequency averaging in the
signal processing as well as effects due to source structure.  
The method is applied  to interferometric visibility data on  
the Vela pulsar which show slightly less scintillation modulation
than expected for a point source in the strong scattering regime.  
The decreased scintillation modulation 
is likely to be due exclusively to time-frequency averaging 
rather than from any source  size effects.  
The implied upper limit on source extent, derived through
 Bayesian inference,  is compared to predictions
that take into account
beaming from the relativistic plasma flow in neutron star 
magnetospheres. 
The upper limit for the transverse source size ($\lesssim 400$ km at
95\% confidence for a pulsar distance of 0.5 kpc)
is easily consistent with conventional
models for radio emission regions in pulsar magnetospheres that place
them well inside the light cylinder at only a few neutron-star radii
from the star's surface.

\end{abstract}

\section{Introduction}
\label{sec:intro}

Diffractive interstellar scintillation (DISS) 
is caused by multipath scattering of radio
waves from small-scale irregularities in the ionized interstellar
medium.  In a previous paper (Cordes 2000; hereafter paper I) methodologies have
been presented that exploit DISS as a superresolution phenomenon: one that can
constrain or determine source sizes on scales that are orders of magnitude smaller
than the inherent diffraction resolution of the largest apertures available, those
involving space VLBI.   In this paper we use some of the methods of Paper I to
assess recent results reported on the Vela pulsar 
(Gwinn \etal~1997, 2000; hereafter \GI~ and \GII, respectively), 
including the conclusion 
that the magnetosphere of that pulsar has been resolved.   

\GI~ analyzed VLBI observations
of the Vela pulsar
using time and frequency resolutions that exploit the spatial
resolving power of  DISS.
Through estimation of the probability density function (PDF) of
the visibility function magnitude, they infer that 
DISS shows  less modulation than expected from a point source
and therefore conclude that the source must be extended.  They 
estimate a transverse size 
$\sim 500$ km for the region responsible for  
the pulsed flux in a narrow range of pulse phase.

We reconsider the observations of G97 because effects
other than source size
can diminish the modulation index (the rms fractional modulation)
from the unity value expected for a point source
in the strong scattering regime
(see Rickett 1990 and Cordes \& Lazio 1991 for definitions of this
regime).   
In particular, even modest averaging in time 
and in frequency over a narrow bandwidth
--- a feature of essentially all observations of DISS ---
can reduce the modulation index to below unity.
We show explicitly that  the averaging parameters used by
\GI~ can account for the entire reduction of the modulation 
index.     To interpret their data, we adopt a Bayesian method to
place limits on the source size. 

\GII~ presented another analysis of data on Vela that took into account
some aspects of time-frequency averaging and other effects that can alter
the visibility PDF.   They revise downward the estimate of source size
to $\sim 440$ km.  Our treatment in this paper contests this result also.
However, the biggest uncertainty in establishing a detection of
or limit on source size is in the DISS parameters, the scintillation
bandwidth and time scale.  We discuss the role of the uncertainties in
these quantities in some detail. 

In \S\ref{sec:vela} we summarize relevant observations of the Vela pulsar
and in 
\S\ref{sec:resolve} we outline the use of DISS 
to resolve sources,  including  a discussion of isoplanatic scales
and formulae for the second moment and the 
PDF of measured interferometer visibilities.    
In \S\ref{sec:bayesian} we apply  a Bayesian inference method
to data on the Vela pulsar which yields an upper bound on 
the source size.
\S\ref{sec:geo} discusses the anticipated 
scale sizes of pulsar emission regions  and interprets our upper limit
on Vela pulsar.  
The paper is summarized in \S\ref{sec:summary}. 
Much of the paper refers to Paper I.

\section{ Observations of the Vela Pulsar}
\label{sec:vela}

The Vela pulsar is ideal for DISS studies that aim at resolving its
magnetosphere because it is very bright and, as shown below, the DISS
isoplanatic scale is smaller than the magnetosphere of the pulsar.
\GI~ report visibility measurements on
the Vela pulsar made at a frequency $\sim 2.5$ GHz.  They show that
the scattering diameter of this strongly scattered pulsar is 
slightly elongated, with dimensions (FWHM) of 
$3.3\pm0.2 \times 2.0\pm 0.1$ mas.  
They also    report values for the DISS time scale and bandwidth of
$\dtd \approx 15$ s and $\dnud \approx 39\pm 7 $ kHz, respectively. 
By comparing the angular diameter with that predicted from the DISS
parameters for specific geometries, they conclude that scattering occurs 
in a thin scattering screen approximately
27\% of the total pulsar-earth distance from the pulsar,
i.e. $\ds/d = 0.27$, where $\ds$ is the pulsar-screen distance and
$d$ is the pulsar-Earth distance.  
The angular size, DISS bandwidth,  and the inferred ratio $\ds/d$ 
differ significantly from those reported by Desai \etal~(1992) 
($1.6\pm 0.2$ mas,  $68\pm 5$ kHz, and 0.81, respectively). 
The DISS bandwidth measured by \GI~ agrees, however, with
that reported by Cordes (1986) when scaled to 2.5 GHz using 
a $\nu^4$ scaling law.
More recently, \GII~ quote different values for the DISS parameters
at $\sim 2.5$ GHz, $\dtd \approx 26\pm1$s and $\dnud \approx 66\pm 1$ kHz.
The exact values of the DISS parameters have a strong influence on the
visibility fluctuations and on any inference from those fluctuations
on source size. 

\begin{figure}
\plotfiddle{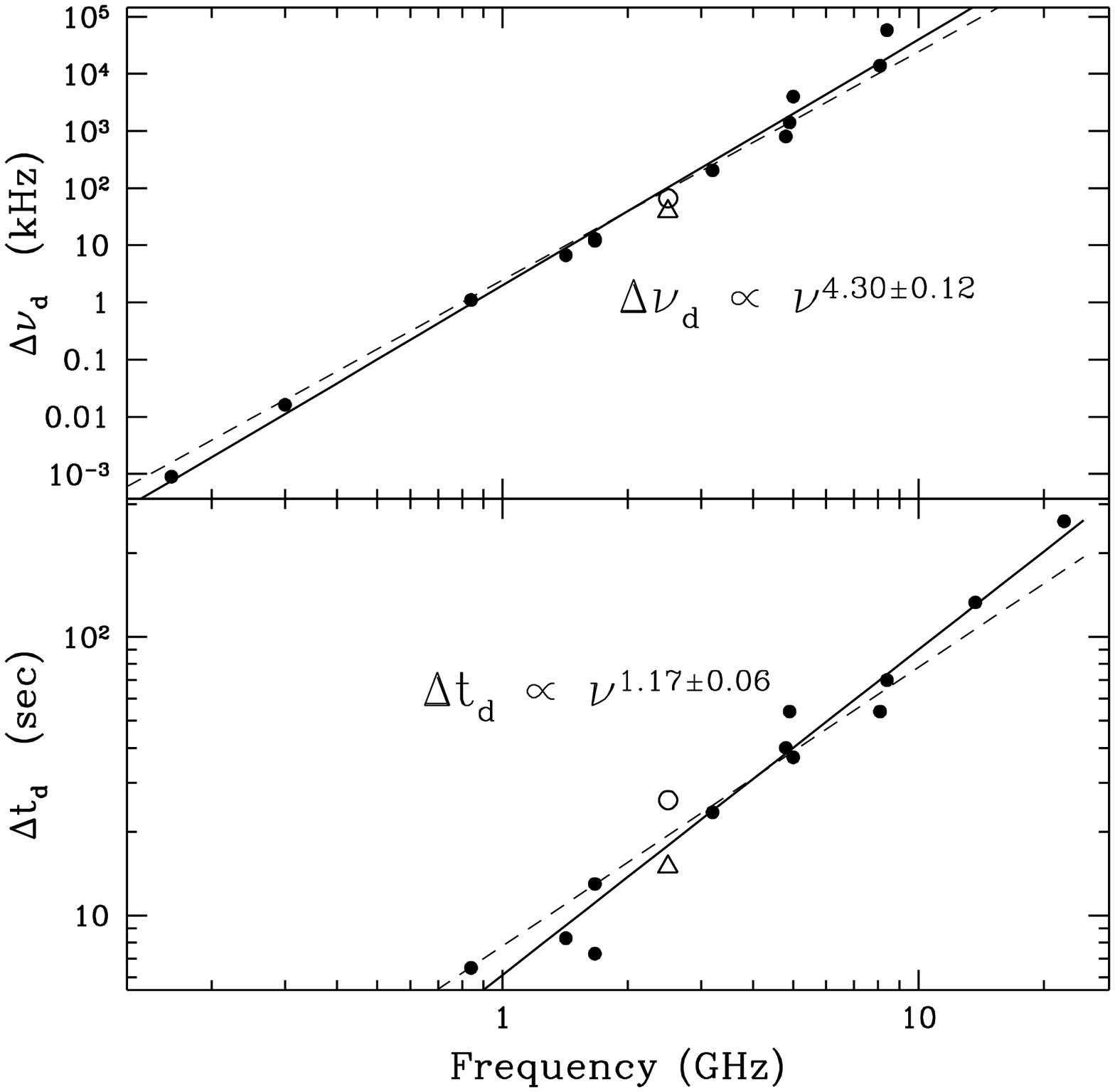}{4.5truein}{0}{70}{70}{-230}{-80}
\caption{ Diffractive scintillation parameters plotted against frequency.
The filled circles are from Johnston \etal~(1998) and references 
therein.  The open triangles are from \GI~ and the open 
circles are values from \GII.    
The solid lines are  unweighted least squares fits to the filled circles
of log-log quantities; errors given are $\pm1\sigma$.  
The best fit scaling laws  are shown.  Also shown
as dashed lines are the scaling laws expected for a square law structure
function, i.e. $\dnud \propto \nu^4$ and $\dtd \propto \nu$.  
(Top Panel:) Scintillation bandwidth, $\Delta \nu_d$.
The exponent is slightly smaller than the value of 
22/5 expected for a Kolmogorov medium.
(Bottom Panel:) Scintillation time scale, $\Delta t_d$.
The exponent  is very close to that expected for a Kolmogorov
medium, which is 6/5.
}
\label{fig:velaiss}
\end{figure}

Some DISS observations on the Vela pulsar 
(Cordes, Weisberg \& Boriakoff 1985; 
Johnston \etal~1998; Backer 1974) 
have suggested that the DISS bandwidth scales
very nearly as $\nu^4$, somewhat different from the $\nu^{22/5}$
scaling commonly associated with the Kolmogorov wavenumber  spectrum
(Rickett 1990).  However,  a composite of all available data from
the literature, shown in Figure~\ref{fig:velaiss}, suggests otherwise.  
Simple, unweighted least-squares fits of log-log quantities
yields $\dnud \propto \nu^{4.30\pm0.12}$, closer to the Kolmogorov scaling
than to the $\nu^4$ scaling.
The increase in value of the exponent is due largely to the new
high frequency measurements by Johnston \etal~1998.  It is conceivable
that there is not a single exponent value for the entire range of
frequencies plotted, but the errors do not allow any break point to
be established. 
Similarly, the DISS time scale varies as $\dtd \propto \nu^{1.17\pm0.06}$,
very close to the $\nu^{6/5}$ Kolmogorov scaling.  Residuals from the fits
imply that there are typically 9\% errors on $\dtd$ and 22\% errors on
$\dnud$.   We use these errors in our inference on source size
rather than using the errors quoted by \GII.   We think the latter
are underestimated.

As shown in Paper I and also as calculated by \GI,
the isoplanatic scale at the pulsar's location can be calculated from
the measured  DISS parameters and scattering diameter. 
The length scale $\ld$ in the diffraction pattern in the observer's plane
is related to the measured angular diameter $\FWHM$ (assuming a
circular, Gaussian brightness distribution and a thin screen) 
and to the DISS bandwidth as
(c.f. Cordes \& Rickett 1998, \S 3.)
\be
\ld 	= \frac	{\lambda \sqrt{2\ln 2}} 
		{\pi \FWHM}	
	= \left[
	    \left ( \frac{c d \dnud}{2\pi\nu^2C_1} \right )
	    \left ( \frac{d}{\ds} - 1 \right )
	   \right]^{1/2}, 	
\label{eq:ld}
\ee
where $C_1=0.957$ for a thin screen with a Kolmogorov spectrum
while $C_1=1$ for a thin screen with a square-law structure function.
Using a nominal value for $\FWHM$,  2.6 mas, taken to be the geometric mean
of the angular diameters measured by Gwinn \etal, we obtain
$\ld \approx 10^{3.6}$ km. 
Solving for $\ds/d$ 
using nominal values for $\FWHM$, the DISS  bandwidth (from \GI), and 
the distance $d=0.5$ kpc,  we obtain $\ds/d \approx 0.27$, in agreement
with \GI.   We note, however, that the true distance to
the pulsar is not well known.  Recent work (Cha, Sembach \& Danks 1999)
suggests that $d$ may be as small as 0.25 kpc, based on association
of the pulsar with the Vela supernova remnant.   This would yield 
$\ds/d \approx 0.15$.  
If we use the DISS bandwidth from \GII, however, we get
$\ds/d \approx 0.39$ for $d=0.5$ kpc and 
$\ds/d \approx 0.24$ using the smaller distance.

\begin{figure}
\plotfiddle{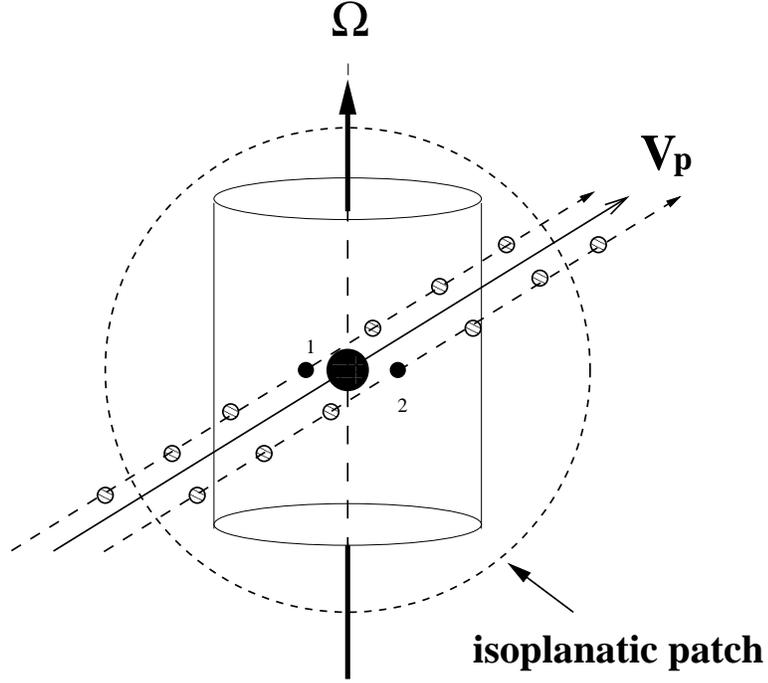}{4.5truein}{0}{70}{70}{-230}{-80}
\caption{
Schematic geometry for a pulsar magnetosphere, showing the light cylinder
of radius $c/\Omega$, the spin axis $\Omega$, and the pulsar's velocity
vector, $\vp$.   The large filled circle denotes the neutron star
while the small circles labelled 1 and 2 represent the locations of
emission regions at the times when radiation is beamed toward Earth to 
produce two pulse components.  The open circles along the straight
dashed lines show the locations of the emission regions as the pulsar
moves translationally.  The pairs of circles are separated by the distance
the pulsar travels in one spin period.   (Not shown is the simultaneous motion
of the light cylinder.) 
The large dashed circle represents the isoplanatic
patch size (not to scale).  
For many cases, the isoplanatic patch is significantly larger
than the light cylinder.  For the Vela pulsar, however, the patch
is smaller than the light cylinder (see text). 
}
\label{fig:geometry}
\end{figure}

In Paper I, we define the isoplanatic scale at the pulsar,
$\drsiso$, as the separation that two sources would have if their 
scintillations are correlated at  a level of $e^{-1}$.  
This scale is related to the diffraction scale $\ld$ as
\be
\drsiso = \frac{(\ds/d)\ld}  {1 - \ds/d}.
\label{eq:ld-drsiso}
\ee
Note that the expression is not valid for $\ds\to d$ because
the small-angle approximation employed in all of our
analysis breaks down.
Using DISS parameters from \GI,
the isoplanatic scale $\drsiso \approx 10^{3.1}$ km for $d=0.5$ kpc
and  $\drsiso \approx 10^{2.8}$ km for half the distance, $d=0.25$ kpc.
Using \GII~ parameters, we get $\drsiso = 10^{3.4}$ and
$10^{3.1}$ km for the two distances.
For comparison,
the light cylinder radius is $\rlc = cP/2\pi\approx 10^{3.6}$ km 
(where the spin period is $P=0.089$ sec),
which is larger than the isoplanatic scale by a factor of 3 to 7.
Figure ~\ref{fig:geometry} shows schematically the pulsar geometry
and its relationship to the isoplanatic scale.  We discuss the geometry
further in \S\ref{sec:geo} below.   Later in the paper, we place an
upper limit on the source size by choosing the largest of all estimates
for the isoplanatic scale, $\drsiso = 10^{3.4}$ km.
This gives the least restrictive upper bound.

\GI~ present a histogram of visibility magnitudes
on a baseline that does not resolve the scattering disk.  
The visibilities 
were calculated using
time-bandwidth averaging 
intervals of  $T = 10$ s and $B = 25$ kHz that resolve the DISS.
They integrated over a 
pulse phase window of 1.16 ms, corresponding to
$\Delta t/P = 0.013$ cycles. 
They also summed over $N_p = 112$ pulse periods to yield their
quoted integration time of 10 sec.  

The reported histogram departs from 
the shape expected for a scintillating point source,
which Gwinn \etal~interpret as 
a signature for source extension, with source size
$\approx 460 \pm 110$ km.   This result is based on the
larger distance estimate for the pulsar (0.5 kpc). 
The quoted error is said to arise
from uncertainty in the scintillation bandwidth, $\dnud$. 

\GI~ did not consider the effects of the time-bandwidth
averaging on their results.   Though $T$ and $B$ are respectively smaller
than the nominal DISS time scale and bandwidth, $\dtd$ and $\dnud$,
they are large enough to be important in any consideration of
visibility fluctuations.   In fact, we show that TB averaging can account
for {\it all} the departure of the fluctuations from those expected for
a point source.  In our analysis below, we also consider contributions to
visibility fluctuations from intrinsic variations in the pulsar.   
We discuss these in some detail in Paper I (Appendix A), where we present
the scintillated, amplitude-modulated noise model.   Amplitude fluctuations
include the well-known pulse-to-pulse variations 
that typically represent 100\% modulations of the pulsed flux at a fixed
pulse phase.

\GII~ presented a new analysis of VLBI data on the Vela pulsar and 
considered TB averaging on their results.   Also, their re-estimated
scintillation parameters reduce the role of TB averaging from what it
would have been using the \GI~ estimates so that their inferred source
size is much the same as in \GI. 
We argue that this new analysis is incorrect, though the DISS parameters
are closer to those extrapolated from other measurements. 
 
\section{Resolving Sources with DISS}\label{sec:resolve}

In Paper I we discuss quantitatively how measurements of interferometric
visibilities (or single-aperture intensities) contain information
about the intrinsic source size, even for sources unresolved
by the interferometer baseline.   The issue is whether the source
is extended enough to reduce the scintillation modulation.
Our discussion is based on the
strong-scattering regime, which easily applies to observations of the
Vela pulsar.   A number of methods may be used to extract source-size
information, including an investigation of the modulation index
of the visibility  and, more accurately, the full probability density
function (PDF) of the visibility magnitude.   For pulsars with multiple
pulse components (and presumably multiple emission regions), cross-correlation
analyses may also be conducted.

\subsection{Second Moments \& Isoplanatic Scales}

To evaluate visibility statistics, we need the autocovariance (ACV) function
$\gamG$
of the `gain' $G$ by which the intensity or visibility is modulated by DISS.
The gain has unit mean, $\langle G \rangle = 1$.
In the strong scattering (Rayleigh) limit   the ACV for a spatial offset
$\drsvec$ at the source, a temporal offset $\tau$ and  an
interferometer baseline $b$ is
\be
\gamG (\bvec, \tau, \delta\nu, \drsvec) = 
\left\langle
G(\rvec, t, \nu, \rsvec)G(\rvec+\bvec, t+\tau, \nu+\delta\nu, \rsvec+\delta\rsvec)
\right\rangle
-1 = \left \vert \gamg(\rvec, t, \nu, \rsvec) \right\vert^2.  
\label{eq:gamG}
\ee
The rightmost equality uses the second moment,
\be
\gamg(\bvec, \tau, \delta\nu, \drsvec) =
\left\langle
g(\rvec, t, \nu, \rsvec)g(\rvec+\bvec, t+\tau, \nu+\delta\nu, \rsvec+\delta\rsvec)
\right\rangle,
\ee
where $g$ is the wave propagator defined in Paper I.
For zero frequency lag $\gamg$ is related to the phase structure function
$\dphi$ by the well-known relation (e.g. Rickett 1990), 
\be
\gamg(\bvec, \tau, \delta\nu=0, \drsvec) =
e^{-\dphi(\bvec, \tau,  \drsvec)}. 
\label{eq:gamg}
\ee
In Paper I we give a general expression for $\dphi$ that applies to
any distribution of scattering material along the line of sight.   For the
Vela pulsar, scattering evidently is dominated by a thin screen along the
line of sight (Desai \etal~1992).   For a thin screen at distance 
$\ds$ from the pulsar, we have (Paper I and references therein) 
\be
\dphi(\bvec, \tau, \drsvec) &\propto& 
		\left ( 
			\frac{\vert \beffvec(s) \vert} {b_e}
		\right )^{\alpha}
   \label{eq:dphi}\\ 
\beffvec(\bvec, \tau, \drsvec)  &=& 
	(\ds/d) \bvec + \veffvec\tau + (1-\ds/d) \drsvec 
   \label{eq:beffvec}\\
\veffvec &=& (\ds/d)\vobsvec + 
	(1-\ds/d)\vpvec - \vismvec(\ds), 
   \label{eq:veffvec}
\ee
where $b_e$ is the characteristic length scale at which $\dphi = 1$ rad$^2$.
A square-law structure function with $\alpha = 2$ may apply to
the Vela pulsar's line of sight, as discussed in \S\ref{sec:vela}, 
though the bulk of the evidence suggests that $\alpha$ is close to 
the Kolmogorov value of $5/3$.
In Eq.~\ref{eq:veffvec}
$\vpvec$ is the pulsar velocity, $\vobsvec$ is the observer's velocity
and $\vismvec$ is the velocity of the scattering material in the ISM.
Note that the offset between sources, $\drsvec$, may be smaller or larger
than the spatial offset associated with the pulsar velocity,
$\vpvec\tau$.

Isoplanatic scales are defined using  $\gamG = e^{-1}$, 
implying $\dphi = 1$ and $\vert \beffvec \vert = b_e$.   
The length scale for the diffraction pattern (at the observer's location)
is given by $\dphi(\ld, 0, 0) = 1$ while the isoplanatic scale at
the source is given by $\dphi(0, 0, \drsiso) = 1$.  Solving for 
$\ld$ and $\drsiso$ and eliminating $b_e$ yields Eq.~\ref{eq:ld-drsiso}.

\subsection{Modulation Index}
\label{sec:mod}

The modulation index of the visibility magnitude, defined as the
rms value divided by the mean intensity, has four terms (c.f. Paper I): 
three associated
with the scintillating source intensity and a fourth due to 
noise, from sky backgrounds and receiver noise, that adds to the
wavefield.   Here we are concerned with
the leading term that is caused by DISS, $\misssq$.
For sources with brightness distribution $I_s(\rsvec)$,  
$\misssq(\bvec, \tau)$ is given by
\be
\misssq(\bvec, \taubar) &=& 
	  \langle I \rangle^{-2}   
		\int\int d\rsvec_1\, d\rsvec_2\, 
		I_s(\rsvec_1)\,
		I_s(\rsvec_2)\, 
		\Qiss(\bvec, \taubar, \rsvec_2 - \rsvec_1, T, B),
\label{eq:miss2}
\ee
where averaging over a time span $T$ and bandwidth $B$ is contained in 
\be
\Qiss(\bvec, \taubar, \delta\rsvec, T, B) 
&=& 
(TB)^{-1}
     	\int_{-T}^{+T} d\tau^{\prime} 
		\left(1 - \left\vert \frac{\tau^{\prime}}{T} \right\vert \right) 
	\int_{-B}^{+B} d\delta\nu\, 
		\left(1 - \left\vert \frac{\delta\nu}{B} \right\vert \right)
\nonumber\\
&~&\quad \quad 
	\times \,
		e^{-ikd^{-1}\bvec\cdot\delta\rsvec} 
	   	  \gamG(0, \tau^{\prime}+\taubar, \delta\nu, \delta\rsvec). 
\label{eq:Qdef}
\ee
In these expressions, $\taubar$ is the time lag that might be imposed in
any data analysis where the visibilities from one site are lagged
with respect to another.  It is easy to show that $\Qiss \le 1$, with equality
only when $T\to 0$, $B\to 0$, and $\delta\rsvec = 0$.   

Inspection of Eq.~\ref{eq:Qdef} shows that 
visibility fluctuations are independent of baseline $\bvec$ for unresolved
sources, for which the complex exponential $\to 1$.  The baseline-independent
property for visibility fluctuations
is similar to the conclusion found by Goodman \& Narayan (1989).  
In general, the integrand factor $\gamG$ in Eq.~\ref{eq:miss2}-\ref{eq:Qdef}
is not factorable.  Practical cases, such as media with 
square-law phase structure functions and Kolmogorov media 
(Paper I), do not show factorability of $\gamG$.  Therefore,
the effects of T-B averaging and source size must be considered
simultaneously. 

Additional contributions to the total modulation index from  
intrinsic source fluctuations and from additive noise are secondary 
to our discussion here.
They are significant, however, in any practical application where 
the time-bandwidth
product is low and where intrinsic source fluctuations are high.
 For pulsars, pulse-to-pulse amplitude
variations are important when only a few pulses are included in any
averaging. 
Appendices B and C of Paper I gives full expressions for all
contributions  to intensity variations.
 
\subsection{Number of Degrees of Freedom in Fluctuations}\label{sec:ndof}

As Eq.~\ref{eq:miss2}-\ref{eq:Qdef} indicate, 
TB averaging and extended structure 
represent  integrals over $\gamG$ that diminish 
scintillation fluctuations.
The modulation index of the averaged intensity or
visibility  depends on
time averaging and source extension in similar ways  because both
increase the number of degrees of freedom 
in the integrated intensity.
The number of degrees of freedom is 
\be 
N_{\rm dof} = 2\misssqinv = 2\niss \ge 2,
\label{eq:ndof}
\ee
where $\niss$ is the number of independent DISS fluctuations 
(``scintles'') that are averaged.  For observations in the 
speckle regime, where scintles are resolved in time and frequency,
we expect $1 \le \niss \lesssim 2$.


If $\misssq = 1$ (within errors), the source is unresolved 
by the DISS, the baseline $\bvec$ has not resolved the scattering
disk, {\it and} the scintillations cannot have decorrelated over
the averaging intervals $T$ and $B$.  The DISS gain $G$
then has an exponential
PDF associated with the two degrees of freedom in the scattered
wavefield. 

Alternatively, $\misssq < 1$ can signify 
(1) variation of the DISS over the averaging time or averaging bandwidth;
{\it or} that
(2) the source has been resolved by the DISS, i.e. that it is 
comparable to or larger than the isoplanatic scale of the 
DISS.  
To discriminate between these possibilities, auxiliary information  
is needed 
that characterizes the dependence of $\gamg$ 
on its four arguments, $\bvec, \tau, \delta\nu$, and $\delta\rsvec$. 
Such information is obtained by making DISS and angular broadening 
measurements over a wide range of frequencies (e.g. Rickett 1990).
As discussed in \S\ref{sec:vela}, evidence suggests that a Kolmogorov 
medium distributed in a thin screen  is relevant.

Complications in estimating $\misssq$ arise from the fact that scintillating
sources fluctuate, on inverse-bandwidth time scales and on a variety of
longer time scales, and there is additive noise in any real-world receiver 
system.  We consider all such complications in Paper I   (main text and 
Appendices B and C).  

\subsection{PDF of Visibility Fluctuations}

\def\Ncal{{\cal N}}
\def\Ncaliave{{\langle {\cal N}_i\rangle}}

The full histogram of visibility fluctuations is potentially much more
sensitive to source structure than is the second moment, as suggested
by Gwinn \etal~(1997, 1998).  Here we summarize our derivation in Paper I of
the PDF that takes into account intrinsic source fluctuations, which are
important for pulsars. 

We assume that the interferometer baseline $\bvec_{ij}$ resolves
neither the source nor the scattering disk.   
Then the time-average visibility can be written as
\be
\Gambar \approx  \Gbar\Iave + \Ncaliave\delta_{ij}  + X + C, 
\label{eq:GambarSAMN}
\ee
where 
$\Iave$ is the mean source intensity,
$\delta_{ij}$ is the Kronecker delta, 
$\langle \Ncal_i \rangle$ is the mean 
background noise intensity,
$X$ is a real Gaussian random variable (RV) with zero mean, and  
$C$ is a complex Gaussian RV with zero mean.
Source fluctuations  are described by $X$ which includes the noise fluctuations
and amplitude fluctuations in the amplitude modulated noise model
(c.f. Appendix C of Paper I). 
$C$ includes additive 
radiometer noise combined with source noise fluctuations, but is uninfluenced
by source amplitude fluctuations.
Expressions for $\sigma_X^2$ and $\sigma_C^2$ are given 
in Appendix C of Paper I.

\def\inorm{{i}}

The PDF for the visibility magnitude is calculated 
by successively integrating over
the PDFs for the different, independent terms in Eq.~\ref{eq:GambarSAMN},
as done in Appendix C of Paper I.
The PDF for the scaled visibility magnitude, 
$\gamma =  \Gambarmag / \sigC$, is 
\be
\fgam(\gamma) &=& 
	\int d\Gbar f_{\Gbar}(\Gbar) 
	\int dX\, f_X(X)
	   \left [
		\gamma e^{-\half (\gamma^2 + \Gbar^2 \inorm^2) } I_0(\gamma \inorm) 
	   \right ]_{\inorm = (\Iave+ X/G)/\sigC}  
\label{eq:fgam1}
\ee
where $I_0$ is the modified Bessel function.  
The integrand factor in square brackets is the 
Rice-Nakagami PDF of  a signal phasor added to complex noise
(e.g. Thompson, Moran \& Swenson 1991, p. 260). 
In the absence of any source, the PDF is simply
$\fgam(\gamma) = \gamma e^{-\gamma^2/2}$.

\subsubsection{Pulsar Noise Contributions}

In Paper I we give detailed examples of the dependence
of the visibility PDF on signal to noise ratio and on the
$X$ and $C$ terms.  In that paper,  Eq. C20 gives the PDF of the complex
visibility and Eq. C25 gives the PDF of the magnitude of the visibility.
Our form for the visibility PDF differs significantly from the equivalent
expression in Eq. 11 of \GI.  The main differences are that
(1) we have unequal variances of the real and imaginary parts of $\Gambar$; 
(2) there is a superfluous factor of $2\pi$ in Eq. 11 of \GI;
and 
(3) the definition of the variances of the real and imaginary parts
in \GI~ is a factor of two too large.

\GI~ argue that a pulsar's contributions to visibility fluctuations
become unimportant for large averaging times.  While the pulsar fluctuations
certainly decrease with increased averaging,
it is also true for the additive radiometer noise, which contributes to
the $C$ term in Eq.~\ref{eq:GambarSAMN}.   Thus, if one is ignored, both
should be ignored.  However, neither should be ignored, as the shape of the PDF
depends on both.   In fact, the ratio of pulsar and radiometer contributions
is independent of the averaging time and depends only on the
system signal to noise ratio, $\Iave /\Ncaliave$.  In practice, it takes
a very strong source for the $X$ term and portions of the $C$ term to
be significant.  The Vela pulsar is marginally strong enough for this to be 
the case.   We find that the PDF shape is altered by pulsar fluctuations
at the level of about 1\% at a radio frequency of 2.5 GHz 
for the Vela pulsar (c.f. Figure 11 of Paper I).

\subsubsection{Sensitivity of the PDF to Time-Bandwidth Averaging} 

As pointed out by \GI~ and \GII, the mode of the visibility PDF
is very sensitive to the number of degrees of freedom in the DISS
fluctuations.  To model the shape of the PDF correctly,  $\niss$
must be correct; but the level of noise fluctuations and the source
intensity must also be correct.  
In paper I we demonstrate how the PDF varies with 
the number of ISS degrees of freedom, $2\niss$.
Increases in $\niss$ can be due to 
time-frequency averaging or source extent or both. 
Statistically, the result is the same.
A pure point source can have statistics
that mimic those given by an extended source if there is sufficient
TB averaging.   Another effect is that if the channel bandwidth $B$ is
varied, $\niss$ is changed but so too is the 
signal to noise ratio.   Decreasing the bandwidth reduces $\niss$
but increases $\sigX$ and $\sigC$ relative to the $G\Iave$ term in
Eq.~\ref{eq:GambarSAMN}, thus masking source-size effects that might also
contribute to the PDF shape.
In addition,  the PDF shape is
sensitive to the nature of the scattering medium
(thin screen vs. extended medium, Kolmogorov vs. square-law structure
function).   For reasons discussed in \S\ref{sec:vela}, we adopt a
thin screen, Kolmogorov scattering medium.  However, we also consider
a thin screen with a square-law structure function.

\section{Bayesian Analysis of Source Size for the Vela Pulsar}
\label{sec:bayesian}

We represent a set of visibility magnitudes
as $\{\vert \Gambar\vert_i, i=1,N \}$, where the bar denotes that
each measurement has been averaged explicitly over time and implicitly
over frequency.  Given a model for the source and parameters
for the scintillations, the likelihood function for
the measurements is (using $\Gambarmag \equiv \gamma\sigma_C$)
\be
{\cal L} = \prod_i \fgam(\gamma_i)/\sigma_C.
\ee
Following a Bayesian scheme presented in Paper I, 
we can infer the posterior PDF for source model parameters $\Thvec$,
\be
f_{\Thvec}(\Thvec) = \frac{{\cal L}}
			{\int d\Thvec {\cal L}}, 
\ee
where we have assumed a flat prior for the model parameters
(see Paper I).

To proceed, we assume that the only source parameter is the
spatial scale $\sigma_r$ of a circular Gaussian brightness distribution,
\be
I_s(\rsvec) = 
	{I_s}_0
	\left ( 2\pi\sigma_r^2 \right )^{-1}
	\exp 	\left ( - \frac {\vert\rsvec \vert^2}  {2\sigma_r^2}
		\right ).
\label{eq:gaussbright}
\ee
We use this distribution to calculate the modulation index (squared)
using Eq.~\ref{eq:miss2} for $\bvec \approx 0, \taubar \approx 0$
and using appropriate values for $T, B, \dtd$ and $\dnud$
(c.f. \S\ref{sec:vela}).   
This yields
$\misssq$ as a function of $\sigma_r$ from which we calculate
the number of ISS fluctuations, $\niss$, using Eq.~\ref{eq:ndof}.
For Kolmogorov media, we have used tabulated values of the
covariance function from Lambert \& Rickett (1999).
This, in turn, we use to calculate the PDF for the DISS gain, $G$,
that is used in Eq.~\ref{eq:fgam1}  to calculate $\fgam$.   Finally,
we calculate $\fGambar$ by appropriate scaling of $\fgam$.

To apply our method, we use the histogram presented in Figure 2 of
\GI, which we write as 
$N_k = N\fGambar(\Gambar_k),\, k = 1, N_{\rm bins}$, 
where $N$ is the total number of visibility measurements
and $N_{\rm bins}$ is the number of bins in the histogram.
Using the histogram, we rewrite the likelihood function as
a product over bins,
\be
{\cal L} = \prod_k \left [\fgam(\gamma_k)/\sigma_C\right]^{N_k}.
\ee


To calculate the PDF for a given source size, we need the parameters
$\SNR \equiv {I_s}_0/\sqrt 2\sigma_+$ (the signal to noise ratio),
$\sigma_+$ (the rms noise $\sigma_C$ when there is no source signal),
and the number of ISS fluctuations summed in each measurement,
$\niss = \niss(T/\dtd, B/\dnud, \sigma_r / \drsiso)$.    
The off-source rms is
$\sigma_+ = \half (\Delta t B N_p)^{-1/2} \langle {\cal N}\rangle$ 
where $\Delta t = 1.12$ ms and $N_p = 112$ s $= T / P$, and $B=25$ kHz.   
In the absence of a source, the rms visibility  is $\sqrt 2\sigma_+$.
We have used  the geometric mean of the off-pulse system noise for the
two sites,
$\langle {\cal N} \rangle  = \left(\Niave \Njave \right)^{1/2}$, expressed
in flux-density units, and the source strength ${I_s}_0$ is the flux 
density in the pulse-phase window used to calculate visibilities.  
This is much larger than the catalogued, period-averaged flux density. 
For Vela, the flux density at 2.5 GHz in the pulse phase window used by
\GI~ is about 10 Jy 
and the intrinsic modulation index of the pulses (from activity in the 
neutron star magnetosphere) is about unity    
(Krishamohan \& Downs 1983).
Evidently, the flux density varies substantially on long time scales
(e.g. Sieber 1973).
This is not surprising given the occurrence of refractive interstellar
scintillations (e.g. Kaspi \& Stinebring 1992).
We therefore
take ${I_s}_0$ to be an unknown and consider \SNR~ to be
a variable to be fitted for. 

\subsection{Analysis of \GI~ Data}

\GI~ report their results in arbitrary correlation units.  We use
the off-pulse PDF in their Figure 2 to estimate (from the mode of the PDF),
$\sigma_+ \approx 550\pm50$ in these units.  We then search over a grid
in $\niss$ and  SNR to maximize the likelihood function.   
The expected mean flux density in the gating window
can be estimated roughly from the shape
of Gwinn \etal's histogram and from theoretical PDFs.  For the 
range of SNR and TB averaging relevant, the mean intensity is 
approximately the number of correlation units where the PDF 
has fallen to 50\% of its peak value.  From Figure 2 of Gwinn \etal,
this is roughly 3600 to 4400 correlation units, implying
SNR $\approx 7.5\pm0.5$.   We use these coarse  estimates solely 
to define a search grid for $\niss$ and \SNR.  

\begin{figure}
\plotfiddle{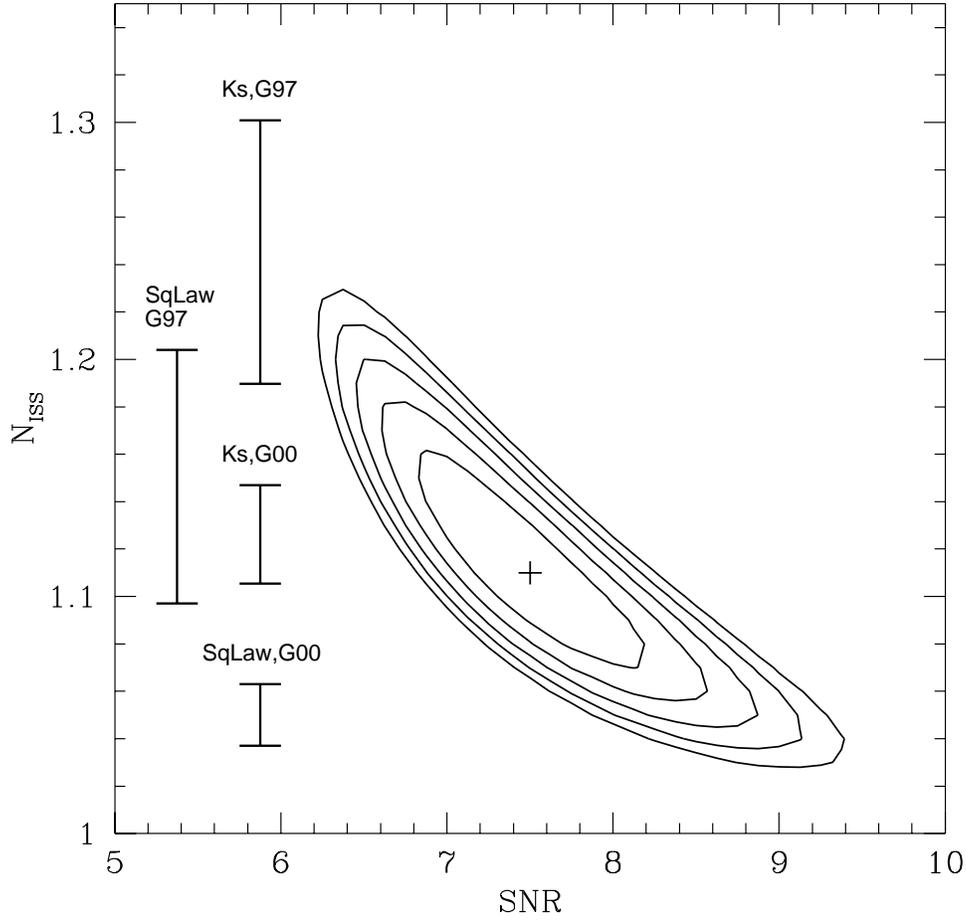}{4.5truein}{0}{65}{65}{-200}{-100}
\caption{
Contours of log likelihood plotted against $\niss$ and 
${\rm SNR}={I_s}_0 / \sqrt{2}\sigma_+$,
respectively the effective number of independent DISS fluctuations
in each visibility  measurement and the signal to noise ratio of 
the pulsar flux density within the pulse-phase
gate.  The contours were calculated  by
maximizing the likelihood for the model visibility PDF 
given the visibility histogram of \GI.
The first contour is at 1/e from the maximum and 
contours are spaced at intervals $\Delta\ln{\cal L} = 1$.
The plus sign indicates the point of maximum likelihood.
The vertical lines with bars indicate ranges of $\niss$
that correspond to different estimates (and errors)
of the DISS parameters (scintillation bandwidth and time scale);
the ranges are calculated by adding $\pm1 \sigma$ to each
of the DISS parameters and are calculated for a point source.
`Ks,G97' refers to DISS parameters quoted by \GI~ and usage
of a Kolmogorov screen to estimate $\niss$.
`SqLaw,G97' uses the same DISS parameters combined with
a screen having a square-law structure function.
`Ks,G00' uses DISS parameters from \GII~ with errors
estimated from our fits in Figure~\ref{fig:velaiss}
and assuming a Kolmogorov screen.
`SqLaw,G00' uses a square-law screen to estimate $\niss$.
The Kolmogorov screen is better supported by the scaling laws
in Figure~\ref{fig:velaiss}.
  We consider the results to signify that
time-bandwidth averaging can account for the best-fit value for
$\niss$ and that the source size is consistent with being point like.  
}
\label{fig:contours}
\end{figure}

Figure~\ref{fig:contours} shows (log) likelihood contours  plotted against
$\niss$ and SNR.  
The plus sign in the figure designates the location of maximum 
likelihood while the vertical lines indicate ranges of
$\niss$ expected using different estimates for the DISS parameters
and using either a scattering screen with a square-law structure function
or a screen with Kolmogorov electron density fluctuations.  
The figure caption gives details.
The best-fit \SNR~ $\approx 7.6$ is consistent with our crude estimate above. 

The contours indicate that $\niss$ and $\SNR$ are anticovariant: 
the best fit 
$\niss$ increases  as $\SNR$ decreases.  This occurs because, as noted
by \GII, the peak of the visibility histogram
is the feature most sensitive to model parameters.  As either $\niss$ or
$\SNR$ increases the peak moves to the right.  
(For example, the variation with $\niss$ is shown in Figure 12 of Paper I.)   
Therefore the two quantities must
compensate each other in order to match the peak  and are thus
negatively correlated.
Based on contours that we do not show, 
we note that smaller values of $\sigma_+$ move the plotted likelihood
contours upward and toward the right.  This also is consistent with the trend
just mentioned.  Therefore acceptable fits are found from a family
of values for $\SNR$, $\niss$ and $\sigma_+$.  

Another source of uncertainty results from the assumed scattering medium. 
The choice of medium doesn't alter the location of the 
contours\footnote{The reason the contours do not change is because
we have used the chi-square PDF with specified $\niss$ to calculate
the PDF of the scintillation gain, $G$.}
but it does alter the mapping of $T$, $B$ and source size to $\niss$.
Usage of a square-law
medium rather than one with Kolmogorov statistics moves 
$\niss$ to lower values..   
Considering uncertainties in the type of medium relevant  
and in the  DISS parameters, 
a conservative conclusion is  that
the maximum likelihood solution
for $\niss$ can be accounted for fully by time-bandwidth averaging 
without invoking any contributions from a finite source size.

\begin{figure}
\plotfiddle{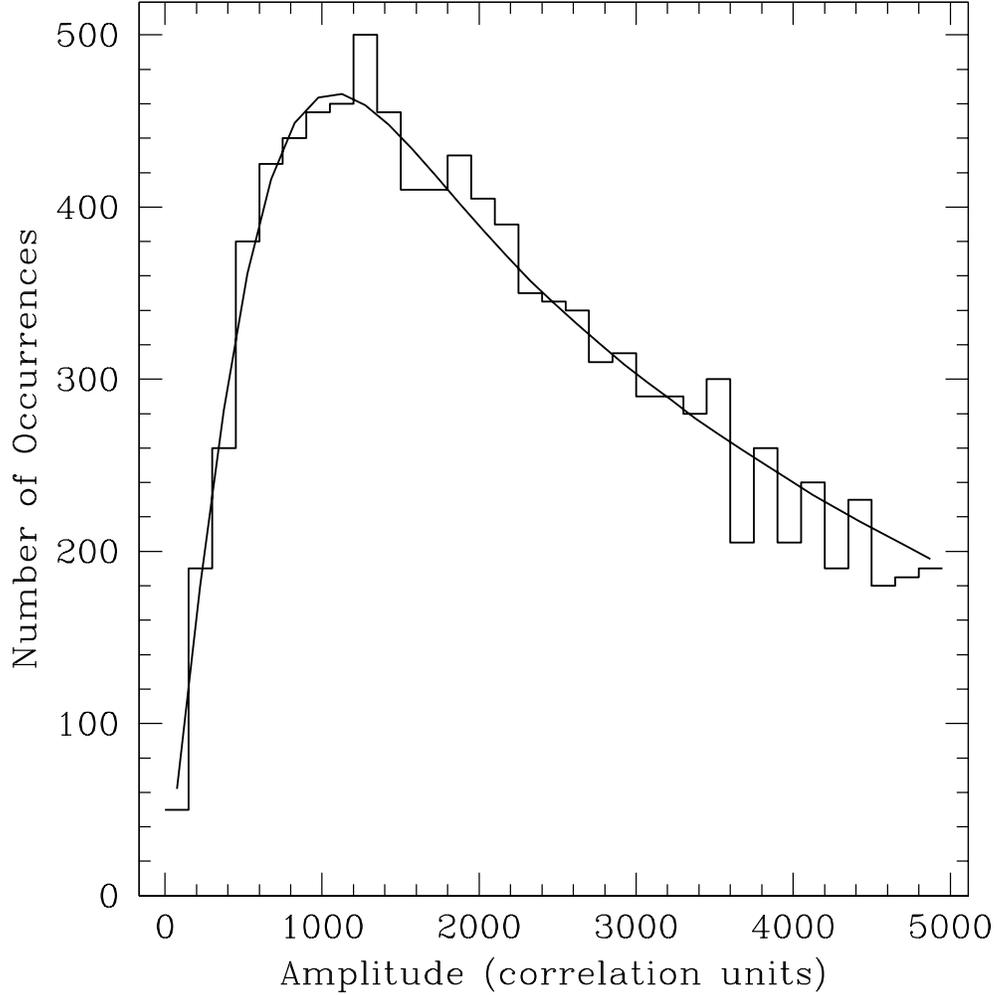}{4.5truein}{0}{70}{70}{-230}{-80}
\caption{
Histogram of measured visibility magnitudes from Gwinn \etal~(1997)
along with our best-fit model (smooth solid line).  As implied by 
Figure ~\ref{fig:contours}, the best fit model is consistent with 
the source being unresolved by the DISS.  
That is, the value for  $\niss$ that yields the best fit
to the histogram  is accounted for
completely by time-frequency averaging in the signal processing.
A finite source size is {\it not} demanded by the data.
}
\label {fig:histogram}
\end{figure}

The observed histogram and the PDF obtained using our best fit parameters
are shown in Figure~\ref{fig:histogram}.    The agreement of the histogram
with our PDF,  
based on a point source and accounting for TB averaging, is as good as
\GI's PDF (their Figure 2),  which was based on an extended source 
but ignored TB averaging.   This shows that not only is our best fit
consistent with $\niss$ expected for a point source (and TB averaging),
but also that it is as good a fit as can be expected.

\begin{figure}
\plotfiddle{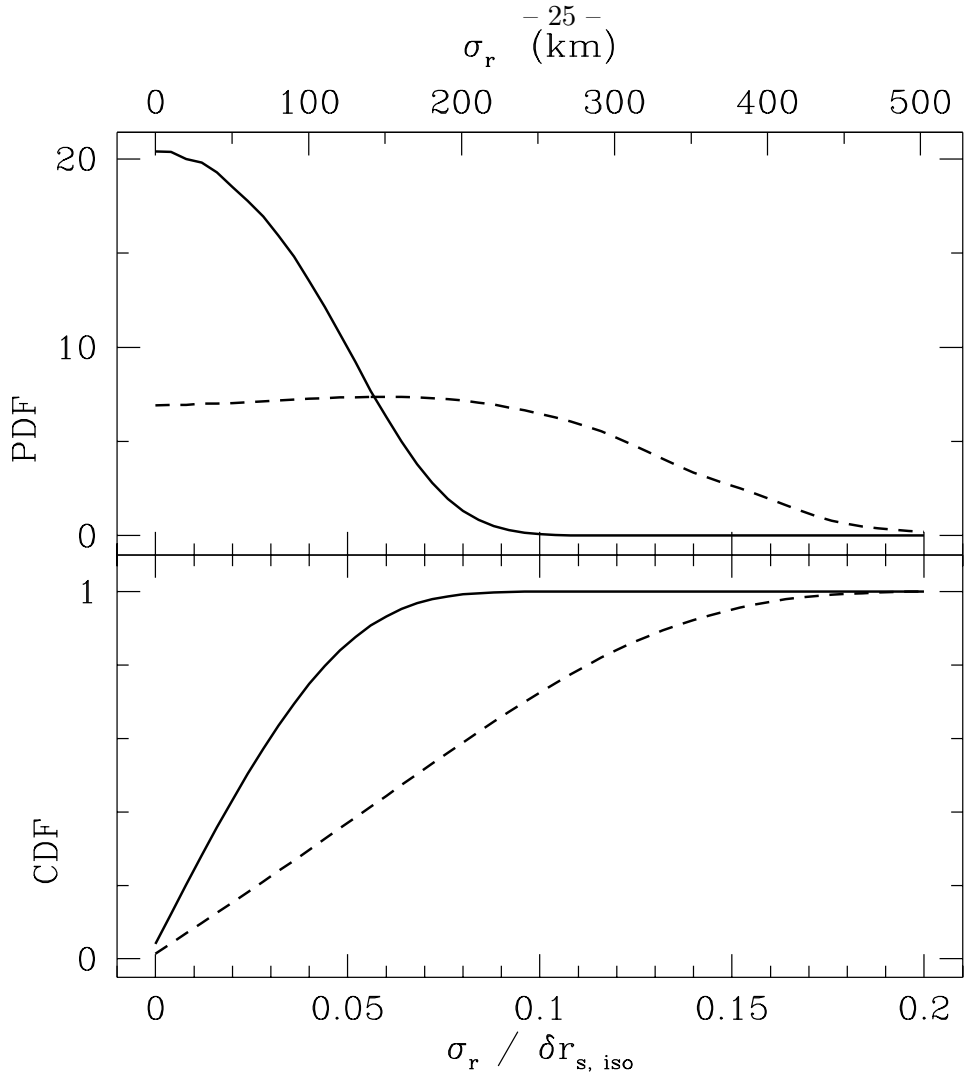}{4.5truein}{0}{70}{70}{-230}{-80}
\caption{
Posterior PDFs (top panel) and CDFs (bottom panel) 
for the source size parameter $\sigma_r$ of a Gaussian
brightness distribution normalized by the isoplanatic scale
$\drsiso$ (bottom scale).  The top horizontal scale expresses
$\sigma_r$ in kilometers, assuming a value for the isoplanatic scale
of $10^{3.4}$ km.  As noted in the text, the isoplanatic scale may
be half this value if the pulsar is 250 pc away rather than 500 pc.
The PDF was calculated by the likelihood function vs.  $\sigma_r/\drsiso$
for fixed values of $\dtd$, $\dnud$, SNR, and $\sigma_+$ and then
marginalizing over these four quantities using distributions
that characterize their uncertainties. 
Solid lines: PDF and CDF calcalated using the best fit values
of SNR and $\sigma_+$ while marginalizing over distributions for 
the DISS parameters.
Dashed lines: PDF and CDF calculated while marginalizing over
distributions for SNR and $\sigma_+$ as well as for the DISS parameters.
}
\label{fig:sourcesizepdfcdf}
\end{figure}

We conclude that the measurements of \GI~ imply only an
upper bound on the source size of the Vela pulsar using the DISS
method. 
To constrain the allowed source size, we recalculate the likelihood by
varying $\sigma_r / \drsiso$, calculating $\misssq$ for a thin-screen,
Kolmogorov medium while accounting for TB averaging  and, hence, calculating
$\niss$ as a function of $\sigma_r/\drsiso$.   
We then use $\niss$ to calculate, in turn, $f_G(G)$, 
the visibility PDF,  the likelihood, and the posterior PDF for
$\sigma_r / \drsiso$.     
These calculations were made by integrating over a grid
of values for $\dtd, \dnud, {\rm SNR}$, and $\sigma_+$
to take into account their uncertainties.
The resultant posterior PDF and CDF for the 
source size are shown in 
Figure~\ref{fig:sourcesizepdfcdf} as dashed lines.
The solid lines are the PDF and CDF obtained when we fix
${\rm SNR}$ and $\sigma_+$ at their best-fit values.
We use the DISS parameter values from \GII~ rather than from \GI~
because we then obtain a larger upper bound on source size. That is,
we derive the least restrictive upper bound.
If we use \GI~ parameter values and a Kolmogorov screen, the
visibility fluctuations are actually quite small compared to
what is predicted.  If we assume a square-law screen, however,
then the $\niss$ value expected from TB averaging is again consistent;
we do not believe, however, that the square-law screen is 
consistent with DISS observations (c.f. Figure~\ref{fig:velaiss}
and previous discussion in text).
In the PDF, the most probable source size is formally  
$\sigma_r /\drsiso \sim 0.04$ but the 
PDF amplitude is nearly the same for zero source size.
We therefore interpret the PDF and CDF in terms of an upper bound
on source size. 
From  the CDF 
we find that $\sigma_r /\drsiso \simless 0.096$
at the 68\% confidence level and 0.16 at 95\% confidence for the case
where we marginalize over all parameters.
Using the {\it largest} of the  isoplanatic scales estimated in 
\S\ref{sec:vela}, 
we obtain an upper limit on source size at 95\% confidence
of 400 km if the pulsar is at $d=0.5$ kpc and
200 km if $d=0.25$ kpc.   Use of alternative values of the DISS
parameters yield even smaller upper bounds.     
These upper bounds are very conservative.  If we fix SNR and $\sigma_+$
at their best fit values, we get an upper bound of 150 km at the 95\%
confidence level.

Clearly, our results imply an upper limit that is potentially 
much smaller than
the source size estimated by \GI. 
The difference is due to our taking into account
the time-bandwidth averaging in the analysis.  To show this, if
we pretend that  
time-bandwidth averaging is negligible and repeat the calculation of
the PDF  for source size, we find a peak that excludes zero source size
corresponding to a size  determination similar to that
of  Gwinn \etal, 
$\sigma_r \approx 0.33\drsiso \approx 415$ km.  
This inference is, of course, erroneous.

\subsection{Analysis of \GII~ Data}

\GII~ present a new analysis of data on the Vela pulsar that takes into
account some aspects of time-bandwidth averaging.   They first fit the visibility
histogram using a finite source size and correct for T-B effects only after
the fact.   Moreover, the scintillation time scale and bandwidth are
estimated simultaneously  in the fitting process, yielding values that
are significantly larger than values presented in \GI.   Larger DISS
parameters cause T-B averaging to be less important  so any apparent
diminuition of the DISS modulation is attributable mostly to 
source size effects.  For these reasons, the source sizes derived
by \GII~ are only slightly less than those found in \GI.  

The analysis in \GII~ appears flawed for the following reasons. 
First of all, the model visibility PDF used in their fits excludes
contributions from pulsar noise.  In practice, this is only about 
a 1\% error in the PDF for the first pulse gate considered by \GII; 
but then, the PDF difference they identify between
a  point-source model and the data is only about 4\% (\GII, Figure 7).
Secondly, they first fit the PDF to find the source size and then,
post facto, correct the source size for time-bandwidth averaging effects.   
Time-bandwidth averaging is handled by calculating one-dimensional
integrals (Equations 6,8 in \GII) and combining them. 
This procedure is not mathematically correct because, as noted in \S\ref{sec:mod},
the scintillation autocovariance  $\gamG$ is not factorable.

To illustrate, consider the fact that during the 10 s averaging time,
the pulsar moves approximately 1400 km given its proper motion of
140 \kms, calculated for a distance of 0.5 kpc (Bailes \etal~1989).   
This is about three times the size inferred by \GII~ for the emission
region in the first pulse gate.  By calculating only a one-dimensional temporal        
integral corresponding to the averaging time and a separate
source-size integral, the combined effects of
source size and averaging are mis-estimated.   This statement follows by
considering the area swept out by the emission region as it is translated
spatially by the pulsar's velocity.   For a circular region of radius
$\sigma_r$, the area swept out $\sim \pi\sigma_r^2(1 + 2\vpperp/\pi\sigma_r)$,
or about three times the area of the emission region itself.  
As shown in Paper I (c.f. Figures 1-6),
the two-dimensional source-size integral causes the scintillation 
variance to decline faster than does the one-dimensional temporal integral. 
The product of the factors is less than the proper multiple integral,
therefore {\it over}estimating $\misssq$ and thus underestimating
$\niss$.

Lastly,  it is unclear exactly how the scintillation parameters
are fitted for in the analysis of \GII.  Apparently they are solved for
while also fitting the visibility histogram
for the size and flux density of a source component.
This differs from the standard procedure of determining the
DISS parameters using the intensity correlation function, so
the systematic errors in the procedure are not known. 
The DISS parameters are quoted
to 1.5\% and 4\% precisions for the scintillation bandwidth and time scale,
respectively.   Given the number of visibility measurements used in the
histogram (Table 2 of \GII) and the implied number of independent ISS
samples used (Eq. 30 of Paper I), the quoted measurement error for the 
scintillation bandwidth seems too small.  Rather, it should be comparable to
the error on the DISS time scale.  Moreover, the scintillation parameters
are significantly larger than presented by \GI.    It is therefore reasonable
to suspect that there are systematic errors on the DISS parameters that
are not included in the quoted uncertainties.    

In Figure \ref{fig:contours} we designate the range of $\niss$ that
is consistent with the DISS parameters quoted by \GII.
For nominal values (26 s and 66 kHz) and assuming a point source,
we obtain $\niss \approx  1.125$ when adopting a Kolmogorov screen.
This value is consistent with the best fit $\niss$ needed to account for the
shape of the histogram.  We therefore conclude,  as before, that the
data of \GI~ are consistent with a point source or, at least, a source
whose size is below
the level of detection in the scintillations.


\def\RGeps{{R_{\Gamma_{\vareps}}}}

\section{Pulsar Geometries}\label{sec:geo}

To interpret the empirical constraints on source size that
we have derived, we now turn to pulsar models and consider
the physical quantities that determine source sizes. 
     

Viable pulsar models associate most radio 
emission with relativistic particle flow along
those magnetic field lines that extend through the velocity of light cylinder.
Emission is beamed in the directions of particles' velocity vectors,
which are combinations of flow along field lines and corotation velocities.
In the following, we assume that 
beaming is predominantly tangential to the field lines, with only
modest corrections from rotational aberration.
The total extent of the emission region in rotational latitude and longitude
is small, so corotation aberration is roughly constant over the emission
region.  We also assume for now that any refraction
of radiation in the magnetosphere is negligible.
Considerable evidence supports the view that radio emission radii
are much less than $\rlc$ 
(Rankin 1990; Cordes 1992 and references therein).

\subsection{Size of the Overall Emission Region}

We distinguish between the extent of the overall emission region 
--- where all pulse components originate --- 
from the instantaneous size responsible for emission seen 
at a specific pulse phase.    The  latter is expected to be 
smaller than the former because of relativistic beaming and
pulsar rotation.  

What do we expect for the transverse extent of the overall
emission regions?
Consider the open field-line region of the pulsar 
to be filled with emitting material.   
The natural length scale is, of course,
the light-cylinder radius, $\rlc$.  
But detailed estimates depend
on relativistic beaming, the magnetic-field topology, and the radial
depth and location of emission regions.  
CWB83 estimate the transverse separation of emission regions
responsible for {\it different} pulse components separated by
pulse phase $\Delta\eta$ to be
\be
\Delta\rs = \third\Delta\eta\sin\alpha_{\mu}\,\rem,
\label{eq:geo1}
\ee  
where $\alpha_{\mu}$ is the angle between the spin and magnetic axes
and $\rem$ is the emission radius. 
Figure \ref{fig:geometry} shows schematically the locations of emission
regions for two pulse components, along with the light cylinder and
the isoplanatic patch.
If we consider a small span 
of pulse phase $\Delta\eta$ 
in a single pulse component (as we consider here), the same equation
(\ref{eq:geo1}) applies.  
This estimate assumes constant-altitude, highly-beamed emission
(with essentially infinite Lorentz factor)
in directions tangential to dipolar magnetic field lines. 

\subsection{Instantaneous Sizes of Emission Regions}

\def\pp{{\eta}}
\def\dpp{{\delta\pp}}
\def\Dpp{{\Delta\pp}}
To predict the transverse extent appropriate for a single pulse component,
we need alternative estimates.  We again consider highly-beamed radiation
along tangents to dipolar magnetic field lines, modified slightly by
rotational aberration.  
For simplicity, we assume that the magnetic axis and the
line of sight are both orthogonal to the spin axis.   
The pulse phase $\pp$ is, for given magnetic polar
angle $\theta$ and radius $r$, 
\be
\eta \approx \pm \frac{3}{2}\theta - \frac{2(r-\rbar)}{\rlc};
\label{eq:geo2}
\ee
the second term accounts for propagation retardation and 
rotational aberration (Phillips 1992) which, together, introduce a time
perturbation $2c^{-1}(r-\rbar)$
(for a nonorthogonal rotator, the two would be replaced
by $[1+\sin\alpha_{\mu}]$). 
The dual signs for $\theta$
account for emission from either
side of the magnetic axis,  and we define a reference radius
$\rbar$ that may be a weak function of frequency 
(Blaskiewicz, Cordes \& Wasserman 1991; Phillips 1992). 

\begin{figure}
\plotfiddle{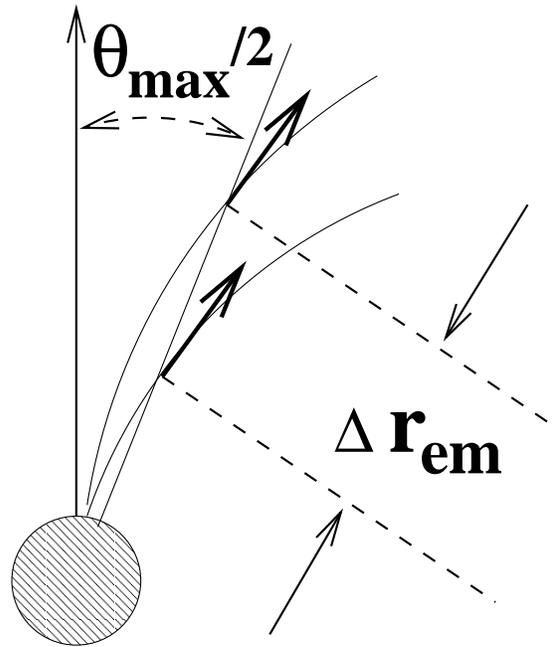}{4.5truein}{0}{90}{90}{-230}{-150}
\caption{
Schematic geometry for relativistic emission from the plasma flow along field
lines near the magnetic pole of a radio pulsar.  The total opening angle
of the radiation region is $\thmax$  and the total radial depth is
$\Delta\rem$.  
}
\label{fig:geometry2}
\end{figure}

It is clear from Eq.~\ref{eq:geo2} 
that different 
$(\theta, r)$ combinations can correspond to the same pulse phase $\eta$.
Thus, a contrived emissivity distribution could produce a very narrow pulse
even for a large radial depth.  
However, we think it more likely that the radial extent $\Drmag$
and the total angular extent $\thmax$ will not be related in this way
and so both will contribute to the overall width of the observed pulse. 
In Figure~\ref{fig:geometry2} we show the geometry and definitions of
the radil depth, $\Drmag$, and the angular width, $\thmax/2$. 
If the radial depth $\Drmag$ is the
same at all $\theta\le\thmax$, then 
\be
\Delta\eta \approx 3\thmax + 2\Drmag/\rlc.
\label{eq:geo3}
\ee

We can place limits on both
$\Drmag$ and $\thmax$ by assuming that either one can dominate the
observed pulse width. 
For an observed pulse width $\Delta\eta/2\pi \approx 0.05$ cycles
(typical of many pulsars, including the Vela pulsar), 
we have the separate limits
\be
\thmax &\lesssim& \third\Delta\eta \label{eq:geo4} \\
\Drmag &\lesssim& \half\Delta\eta \,\rlc. 
\label{eq:geo5}
\ee

Another contribution to the observed pulse width may come from the
radiation beam width of an individual particle. 
If we relax our assumption of highly beamed radiation, 
and allow the pulse width to be determined by the relativistic beam width,
then Lorentz factors
\be
\gamma_p \gtrsim \frac{3}{2\Delta\eta}
\label{eq:geo6}
\ee
are allowed. 

We now wish to calculate the instantaneous source size  $\drs$ relevant
to an observation over a vanishingly small range of pulse phase. 
At a fixed pulse phase, the transverse extent of the emission region
is $\drs = 0$ as $\Drmag\to 0$ and $\gamma_p \to\infty$.
Over a small range of pulse phase, the transverse extent is given by
Eq.~\ref{eq:geo1}.  For finite depth $\Drmag$, the transverse extent
(again for $\gamma_p \to \infty$) is
\be
\drs \approx \half \Drmagperp \approx \Drmag\langle \theta\rangle,
\label{eq:geo7}
\ee
where $\langle\theta\rangle$ is the mean $\theta$ over those
$(r,\theta)$ in Eq.~\ref{eq:geo2} 
 that satisfy $\eta =$ constant.  Taking
$\thmax$ as an upper limit on $\langle \theta\rangle$,
we obtain
\be
\drs \lesssim \half \Drmag\thmax.
\label{eq:geo8}
\ee
If relativistic beaming is finite, the transverse extent due
to finite Lorentz factor $\gamma_p$ is
\be
\drs \lesssim \rem \gamma_p^{-1}. 
\label{eq:geo9}
\ee
Expressing these results in terms of $\rlc$ we have
\be
\frac{\drs}{\rlc} \approx 
	\left\{
	\begin{array}{ll} 
		\half
		\epsem\thmax
		\left(\frac{\Drmag}{\rem}\right)
		\lesssim \frac{1}{12}\left(\Dpp\right)^2
		& \mbox{{\rm finite depth}} \label{eq:geo10} \\
\\
		\epsem\gamma_p^{-1}
		\lesssim \frac{2}{3}\epsem \Dpp 
		& \mbox{{\rm  finite Lorentz factor}},
		\label{eq:geo11}
	\end{array}
	\right.
\ee
where $\epsem \equiv \rem/\rlc$ and
 the limits are based on Eq.~\ref{eq:geo4}-\ref{eq:geo5}.

Pulsar phenomenology suggests that $\epsem \lesssim 0.1$
and $\Dpp/2\pi \lesssim 0.1$ cycles, implying nearly 
identical upper limits in 
Eq.~\ref{eq:geo10}
that are $\sim 3$\% of $\rlc$.  However, the true limits are probably
much smaller.  Rankin (1990, 1993) has shown that pulse widths
(in cycles) scale $\propto P^{-1/2}$, as expected if they 
are determined by the angular extent of the open field-line
region near the magnetic axis.  (A slightly different scaling
has been described by Lyne \& Manchester [1988].)   This implies that any 
contributions from relativistic beaming and radial depth
are small.   If so, the upper limits in 
Eq.~\ref{eq:geo4}-\ref{eq:geo5} are probably a factor of
ten smaller and the lower limit on $\gamma_p$
in Eq.~\ref{eq:geo6} is a factor of 10 larger. 


\subsection{Length Scales in the Vela Pulsar's Magnetosphere}

Using the pulse width (4.5 ms at 10\%) and
period (89 ms) of the Vela pulsar, we have
$\Dpp/2\pi \approx 0.05$ cycles  and Eq.~\ref{eq:geo11} 
yields  upper bounds
on $\drs/\rlc$ of 0.008 and $0.02(\epsem/0.1)$ for radial
depth and Lorentz-factor limits, respectively.  These correspond
to  34 and 89 km (for $\epsem = 0.1)$.   
Rankin (1990) has classified the radio
pulse for the Vela pulsar  as a `core' component whose width is 
consistent with
the angular size of the open field-line region expected
for a dipolar field and for an emission radius
$\rem \lesssim 2\rns$, where $\rns=10$ km is the
assumed neutron star radius.  By contrast, Krisnamohan \& Downs
define four separate components in the radio emission.  
Thus $\rem \lesssim 20$ km,
corresponding to $\epsem \lesssim 10^{-2.3}$.  With 
$\Drmag \lesssim \rem$, we obtain limits
$\drs/\rlc \lesssim 10^{-3.61}$ and $10^{-3.0}$, or
$\drs \approx 1$ and 4 km, respectively.   
The largest relevant transverse extent may in fact derive 
from the finite gating window used by \GI~ and \GII,
which is $\Dpp/2\pi = 0.013$ cycles, yielding 
\be
\drs \approx \third \Dpp\, \rem \approx 118\epsem \,\, {\rm km}.   
\ee
Simple pulsar geometries therefore suggest that the transverse
scales of pulsar emission region(s) in the Vela pulsar
should be less than
the upper bound from the DISS observations and, in fact,
may be very much less.  In addition, using Eq.~\ref{eq:geo6}
and $\Delta\eta/2\pi \approx 0.01$ cycle for the component on the
leading edge of Vela's radio pulse (Krishnamohan \& Downs 1983),
a lower bound on the Lorentz factor is $\gamma_p \gtrsim 150$.

More complicated geometries may ensue if refraction within
the pulsar magnetosphere is important, as has been suggested
by a number of authors 
(Melrose 1979; Barnard \& Arons 1986; Cordes \& Wolszccan 1988; 
\GI; \GII).   Refraction can duct
wavemodes along magnetic field lines and then convert
energy to propagating electromagnetic waves at some altitude
that effectively could be defined as the `emission'
altitude.    Any such ducting may alter
the scalings of radio beam width with period from
those expected from the Goldreich-Julian polar-cap size
and in the absence of refraction.    Differential
refraction might cause radiation from disparate regions of
the magnetosphere to reach the observer simultaneously.
However, like the simple geometry considered here, if refraction is
significant,  one
might also expect pulse widths to be much larger than observed,
or at least larger than predicted from the size of the open-field-line
region. 
That statement,
along with the remarkable consistency of scaling laws for
pulse widths with the Goldreich-Julian polar cap size
and relatively small emission radii, suggests that refraction
does not enlargen the transverse scales from which
radio radiation emerges from pulsar magnetospheres.

We conclude that the magnetosphere of the Vela pulsar is
likely to have radio emission regions with transverse scales too small
to have been detected in the observations of Gwinn \etal~
It is also possible that they will {\it never} be detected using DISS
or any other technique, for that matter, short of traveling to the vicinity
of the pulsar.  

\subsection{Comparison with Gamma-ray Emission}

High-energy emission from the Vela pulsar, such as the $> 100$ MeV
pulsed gamma-rays seen with EGRET, shows a double pulse that is offset
to later pulse phases from the radio pulse.   Rankin (1990) classifies
the radio pulse as a core component which, if similar to core emission
from other pulsars, is consistent with an emission altitude close to
the neutron star surface.   Gamma-ray emission may be high-altitude
``polar-cap'' radiation that derives from the flow along open field
lines originating near the magnetic polar cap (e.g. Harding \& Muslimov 1998).
Alternatively, it may be ``outer-gap'' emission from near the light cylinder
whose beaming is highly affected by rotation (e.g. Romani 1996).
Our conclusion that the radio emission's transverse extent is small, at least in
a narrow range of pulse phase, is consistent with both
of these pictures of $\gamma$-ray emission.     

\section{Summary and Conclusions}\label{sec:summary}

In this paper we applied our superresolution methodology to
the recent VLBI observations of the Vela pulsar by Gwinn \etal~(1997, 2000)
and find that the scintillation statistics may be accounted for fully
by time-bandwidth averaging and a finite signal-to-noise ratio.
  Any contribution from extended source
structure is less than an upper limit of about  400 km at the 95\% confidence
interval.  This limit is larger      
than the size expected from conventional models that place radio
emission well within the light cylinder of the pulsar and therefore is
not restrictive on those models.
Scintillation observations at frequencies lower than 2.5 GHz
have better resolution ---  because the isoplanatic scale
at the pulsar's position scales as $\nu^{1.2}$ --- and thus may 
be able to detect the finite source size.
Recent work by Macquart \etal~ (2000) failed to find any effects of
finite source size at 0.66 GHz.

Our results differ  from those of Gwinn \etal~ for several reasons.
We have used a more complete signal model
combined with a rigorous treatment of time-bandwidth averaging
and consideration of the substantial uncertainties in the measured
scintillation parameters.   These parameters, the diffraction time scale
and bandwidth, determine the number of degrees of freedom 
encompassed by scintillation fluctuations when time-bandwidth averaging
and source-size effects are considered.  This number, in turn, 
has a strong influence on the shape of the visibility histogram,
from which the source size is inferred or constrained.   Another source
of uncertainty is the nature of the scattering medium, viz. the
wavenumber spectrum and phase structure function.  We have found that
the medium is close to being Kolmogorov in form  by investigating the
scaling laws of the scintillation parameters with frequency.
We note also that if the shape of the visibility histogram 
is analyzed incorrectly, the inferred source size will have 
a systematic error that will scale with the isoplanatic length scale
referenced to the location of the source (Eq.~\ref{eq:ld-drsiso}).
The isoplanatic scale, $\drsiso$, scales with frequency approximately
as $\nu^{1.2}$, so one would expect the mis-estimated source size
to scale in the same way.

Our methodology can be applied to any radio source in the strong 
scattering regime,
including compact active galactic nuclei and gamma-ray burst afterglows,
though it is not clear that these sources are compact enough to show
diffractive scintillations in this regime.
In another paper, we will address sources of these types and we will also
consider scintillations in the weak and transition scattering regimes.

I thank Z. Arzoumanian, S. Chatterjee, C. R. Gwinn, H. Lambert, 
T. J. W. Lazio, 
M. McLaughlin,  
and B. J. Rickett for useful discussions
and H. Lambert and B. J. Rickett for making available their 
numerically-derived autocovariance functions for Kolmogorov media. 
This research was supported by NSF grant 9819931 to Cornell
University and by NAIC, which is managed by Cornell University
under a cooperative agreement with the NSF.

\end{document}